# Bistable topological insulator with exciton-polaritons

Yaroslav V. Kartashov[1,2,3] and Dmitry V. Skryabin[1,4]

[1]*Department of Physics, University of Bath, BA2 7AY, Bath, United Kingdom*
[2]*ICFO-Institut de Ciencies Fotoniques, The Barcelona Institute of Science and Technology, 08860 Castelldefels (Barcelona), Spain*
[3]*Institute of Spectroscopy, Russian Academy of Sciences, Troitsk, Moscow Region, 142190, Russia*
[4]*Department of Nanophotonics and Metamaterials, ITMO University, St. Petersburg 197101, Russia*
*Corresponding authors: Yaroslav.Kartashov@icfo.eu; d.v.skryabin@bath.ac.uk*

**Functionality of many nonlinear and quantum optical devices relies on the effect of optical bistability. Using the microcavity exciton-polaritons in a honeycomb arrangement of microcavity pillars, we report the resonance response and bistability of the topological edge states. A balance between the pump, loss and nonlinearity ensures a broad range of dynamical stability and controls the distribution of power between counter-propagating states on the opposite edges of the honeycomb lattice stripe. Tuning energy and polarization of the pump photons, while keeping their momentum constant, we demonstrate control of the propagation direction of the dominant edge state. Our results facilitate development of new applications of topological photonics in practical devices.**

PACS numbers: 71.36.+c, 42.65.Yj, 63.20.kk, 73.21.Cd

The phenomenon of topological insulation and existence of the associated edge states were first encountered and explained in the electronic systems [1]. Nowadays these ideas find application in such diverse areas as atomic physics, matter waves, optics, and acoustics. Topological insulators possess a forbidden energy gap in the bulk, where no states can exist, but when placed in contact with a material having different topological invariants (Chern numbers) [1], they support the in-gap unidirectional and topologically protected from scattering edge states propagating along the interface. One of the prime near future applications of the topological photonic and atomic states is perceived as creation of topologically protected quantum information processing and transmission devices [2-4].

In photonics, topological edge states have been proposed theoretically and recently observed experimentally in many systems, see, e.g., [5-19]. In particular, we would like to mention here arrays of coupled resonators [7,8], including the ones with parametric nonlinearity [14], and the exciton-polariton microcavities, where strong photon-exciton coupling leads to the formation of the half-light half-matter exciton-polariton quasi-particles [15-17]. It is well known that an interplay of nonlinearity, loss, and resonant pumping in microcavities leads to a rich variety of dynamical effects, most of which are underpinned by the existence of a pump frequency range where the intracavity intensity becomes bistable [20]. Majority of theoretical and experimental works on photonic topological insulators has so far dealt with the linear regime of light-matter interaction, but the interest in nonlinear [13,14,18,19,21-24] and associated quantum [14] effects is quickly gaining its pace. In this regard arrays of nonlinear twisted waveguides [18,22,23] and nonlinear exciton-polariton devices [19,21,24] hold strongest promise for the near future experimental realization. Under a variety of conditions nonlinear topological edge states in photonic settings have been found to decay into trains of the edge quasi-solitons [21-24]. Nonlinear effects and solitons in topological insulators are also emerging in a variety of non-photonic system and the whole area is quickly acquiring an interdisciplinary flavor [25-28].

Losses exist in any photonic system and they can complicate observation and practical use of the topological states [29]. When gain is present, then the threshold for topological edge states may be lower than for other available states providing favorable conditions for topological lasing [30-33]. However, when pump is taken as an external resonant forcing a bistable regime of operation of topological photonic insulators may be expected, which has not been so far predicted and demonstrated. Such resonant pumping imprints both energy and momenta of the external laser photons onto the microcavity polaritons. This raises a question about the mere existence of the topological states under such conditions [29,34].

Exciton-polariton condensates represent a very practical example of dissipative non-equilibrium system [35], where bistability is well established, see, e.g. [36], and a variety of periodic potentials [37-39] have been used for observations of one- and two-dimensional polariton lattice solitons [40-42] and linear non-topological edge states [43,44]. Moreover, polaritons are quasi-particles with spin, see, e.g., [35,45], that demonstrate polarization-dependent tunneling between the lattice sites [15,21,46,47] originating in the fact that polaritons in a homogeneous planar resonator have momentum-dependent linear coupling between the states with positive and negative spins, which is formally analogous to the spin-orbit coupling in atomic physics [35,48,49]. Spin-orbit coupling is the key effect allowing realization of the polaritonic topological insulators, see, e.g., [15,16,21,24,34]. For all these reasons an exciton-polariton platform appears to be a natural choice for studying the properties of the topological insulator state in the bistability regime.

The aim of this work is to demonstrate that topological edge states persist in the presence of the resonant external pump, loss, and nonlinear self- and cross-spin interactions. These states can be resonantly excited when the pump photon energy approaches the edge state energy of the loss- and pump-free system. We found that for sufficiently strong pump values the resonance response curve tilts enough in frequency to induce the bistable response of the topological insulator and that polarization of the pump has a profound impact on the power balance between the edge states on the opposite sides of the lattice. We present first stable nonlinear topological edge states and illustrate their robustness upon interaction with structural lattice defects.

We describe the evolution of spinor polariton condensate in a lattice of microcavity pillars, akin to those fabricated in Refs. [46,47], using a system of coupled Gross-Pitaevskii equations for the spin-positive and spin-negative components of the polariton wavefunction $\Psi = (\psi_+, \psi_-)^T$ [24,35]:

$$i\hbar \frac{\partial \psi_\pm}{\partial T} = -\frac{\hbar^2}{2m}\left(\frac{\partial^2}{\partial X^2} + \frac{\partial^2}{\partial Y^2}\right)\psi_\pm + \frac{\beta\hbar^2}{m}\left(\frac{\partial}{\partial X} \mp i\frac{\partial}{\partial Y}\right)^2 \psi_\mp +$$
$$\varepsilon_0 [\mathcal{R}(X,Y)\psi_\pm \pm \Omega\psi_\pm + (|\psi_\pm|^2 + \sigma|\psi_\mp|^2)\psi_\pm - i\gamma\psi_\pm + \mathcal{H}_\pm(Y,T)]. \quad (1)$$

Here $m$ is the polariton mass, which is assumed $10^{-31}$ g; $\beta = 0.3$ is the dimensionless parameter characterizing strength of the spin-orbit coupling [24,49]; $\varepsilon_0$ is the scaling coefficient having the dimension of energy, which characterizes the nonlinear shift of the polariton energy level $\varepsilon_0|\psi_\pm|^2$, relative to the linear resonance; $X, Y$ and $T$ are the physical distances and time, which are replaced by the dimensionless coordinates $(x,y) = (X/L, Y/L)$ and $t = T\varepsilon_0/\hbar$ in our modelling. Here $L = 1~\mu$m, $\varepsilon_0 = \hbar^2/mL^2 \simeq 0.3$ meV and $\hbar/\varepsilon_0 \simeq 2$ ps. Potential energy landscape felt by polaritons is represented by $\mathcal{R}(x,y)$, which is assumed to be a honeycomb lattice of micropillars with a single pillar described by the Gaussian function $-pe^{-[(x-x_m)^2 + (y-y_n)^2]/d^2}$, $d = 0.5$, $p = 8$. The pillars are separated by the dimensionless distance $a = 1.4$. The lattice is truncated along the $x$-axis, and it is kept periodic along $y$ with period $3^{1/2}a$ (Fig. 1). We use truncation that creates so called *zigzag* edges, which are considered throughout this work. Dimensionless $\Omega = 0.5$ is proportional to the applied magnetic field resulting in the Zeeman energy splitting for the two spin states. We also assume that polaritons with the same spin repel, while polaritons with opposite spins weakly attract, $\sigma = -0.05$ [36]. We consider resonant excitation of the microcavity modes by the two-component pump $\mathcal{H}_\pm(y,t) = h_\pm e^{iky-i\varepsilon t}$, where $h_\pm$ are the dimensionless amplitudes, $k$ and $\varepsilon$ are the dimensionless momentum along the edge and normalized pump frequency detuning (photon energy) from the polariton resonance at zero momentum, respectively. The latter also serves as a reference energy in this formulation. $\gamma = 0.01$ and $\hbar/2\gamma\varepsilon_0 \simeq 100$ ps is the polariton lifetime [50]. Results presented below are robust with respect to the realistic changes of the parameter values and, in particular, the bistability effect can also be found for shorter polariton lifetimes encountered in poor quality samples.

We first introduce energy spectrum and band structure of the linear pump- and loss-free system [24]. We seek for linear Bloch waves using a substitution $\psi_\pm(x,y) = u_\pm(x,y)e^{iky-i\varepsilon t}$, where $u_\pm$ are periodic in $y$ with a period $3^{1/2}a$ and localized in $x$, and $\varepsilon$ is a periodic function of the momentum $k$ with a period $K = 2\pi/3^{1/2}a$. We assume that the lattice is truncated at two points along $x$, so that there are two zigzag edges. The time reversal symmetry $(\psi_+, \psi_-, t) \to (\psi_-^*, \psi_+^*, -t)$ violation by $\Omega \neq 0$ is a prerequisite for appearance of unidirectional edge states [5]. A typical $\varepsilon(k)$ dependence around the lowest energy gap and two in-gap topological edge states are shown in Fig. 1(a) [24]. Remnants of Dirac points existing in a 2D lattice are seen in the band part of the spectrum (black lines) in the proximities of $k = K/3$ and $k = 2K/3$. Energies of the in-gap topological edge states located at the left and right edges of the stripe are shown in red and green, respectively. Dots mark momenta where $\varepsilon'' = \partial^2\varepsilon/\partial k^2$ and, respectively, the polariton mass associated with the edge modes change their signs. The mass is positive in the upper half of the gap and negative in the lower one. $\psi_-$ component of the edge states has a larger amplitude than $\psi_+$ in this gap providing $\Omega > 0$. Qualitatively, the polariton spin-orbit coupling leads to appearance of a vortex with charge -2 in the weak $\psi_+$ component when strong $\psi_-$ has the trivial phase. These vortices appear in each potential well and split into pairs of charge -1 vortices due to perturbations, e.g., proximity of the neighboring wells. The phase accumulation direction for negatively charged vortices determines preferential direction of the edge state flow along the interface, which is for this reason has to be opposite at two interfaces, see Fig. 1(b). Change of the magnetic field direction (sign of $\Omega$) changes the dominant spin component and hence the net vorticity reverses its direction leading to the reversal of the propagation directions of the edge states.

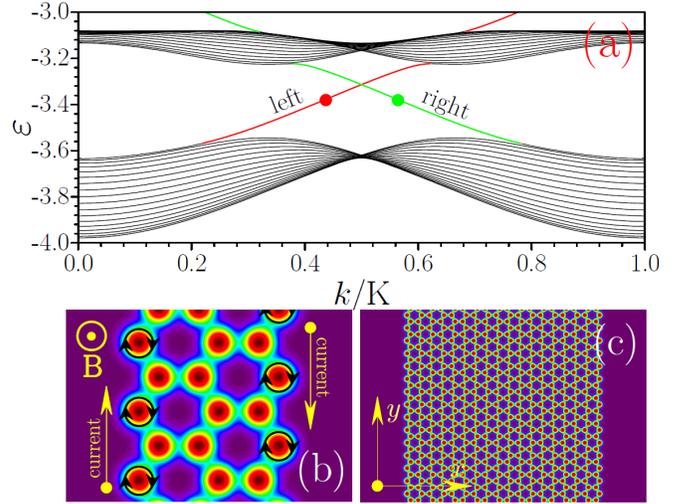

Fig. 1. (Color online) (a) Energy of linear modes versus Bloch momentum for the lattice with two zigzag edges. Black lines correspond to the bulk modes, red and green lines to the edge states. A boundary at which an edge state resides is indicated. Dots mark the momenta, where the effective polariton mass changes its sign. (b) Schematic illustration showing lattice of microcavity pillars periodic in the vertical $y$-direction, black circles with arrows show direction of rotation in vortices induced in the $\psi_+$ component, while yellow arrows show currents in topological edge states that are opposite at opposite edges. Magnetic field $B$ is also indicated. (c) Full-scale lattice that was used in simulations.

We anticipate that the pump $\mathcal{H}_\pm(y,t) = h_\pm e^{iky-i\varepsilon t}$ should resonantly excite the Bloch states of the linear system having energy and momentum close to those of the pump and well isolated from the continuum. Since there exist two edge states for a given momentum $k$, two resonances should in general show up within the gap. To confirm the above conjectures and study the impact of nonlinearity we seek nonlinear modes of the full system (1) in the form $\psi_\pm(x,y) = u_\pm(x,y)e^{iky-i\varepsilon t}$, where $u_\pm$ obey

$$\frac{1}{2}(\partial_x^2 + \partial_y^2 + 2ik\partial_y - k^2)u_\pm - \beta(\partial_x \mp i\partial_y \pm k)^2 u_\mp - \quad (2)$$
$$\mathcal{R} u_\pm \mp \Omega u_\pm - (|u_\pm|^2 + \sigma|u_\mp|^2)u_\pm + i\gamma u_\pm - h_\pm + \varepsilon u_\pm = 0.$$

We solved system (2) with a variant of the Newton method applied in the momentum space using 201 and 25 spatial Fourier harmonics along $x$ and $y$, respectively.

First, we discuss results for the linearly polarized pump field, $h_+ = h_-$, see Fig. 2. We fix the pump momentum to $k = 0.4$ K and scan the pump energy within the gap. The energy of the left edge state in pump-free system for this value of momentum is located in the middle of the gap, while the energy of the right edge state is very close to the continuous spectrum, see Fig. 1(a) and the vertical lines in Figs. 2 (a,b). We found that the former state is resonantly excited when the pump energy matches the resonance of the unforced system, while the latter state does not make a noteworthy response. Typical resonance dependencies of the amplitude of the left edge state on energy detuning $\varepsilon$ are shown in Fig. 2(a) and the corresponding examples of the transverse distributions of the polariton densities are shown in the right column of the same figure. When pump is strong enough, the resonance curve acquires a pronounced nonlinearity-induced tilt and forms a typical bistable loop, see Fig. 2(a).

Figure 2(b) shows an average width $w_x$ of the edge state calculated as per [51]. One can see that the best edge localization is achieved close to the resonance and roughly the same degree of localization is preserved throughout the bistability interval of the energy values. Thus, the degree of localization of the edge states in the externally forced polaritonic topological insulator can be controlled by varying detuning $\varepsilon$. Right column of Fig. 2 also shows transverse maps of the local polarization degree, $\rho = (|\psi_+|^2 - |\psi_-|^2)/(|\psi_+|^2 + |\psi_-|^2)$, associated with the edge sates. The negative polarization (blue colors) dominates inside the pillars located on the edge, but the positive polarization takes over deeper inside the lattice within the bistability interval.

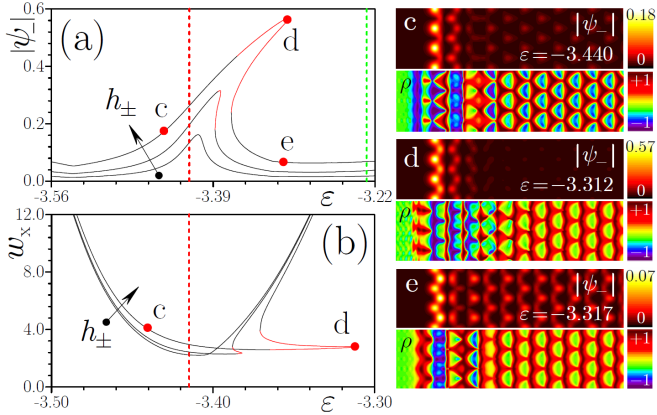

Fig. 2. (Color online) Peak amplitude of spin-negative component (a) and x-width of the edge state (b) vs energy $\varepsilon$ for the pump amplitude $h_\pm = 0.001$, $0.002$, and $0.004$ at $k = 0.4\,\mathrm{K}$. Arrows indicate the direction of the increase of pump amplitude. Red dots correspond to profiles (modulus $|\psi_-|$ and local polarization degree $\rho$) shown in the right column. Stable/unstable states are shown black/red in (a) and (b). Vertical red and green lines mark energies of linear pump-free edge states.

Polarization of the pump is another control parameter that can be used to manipulate the resonance response of the edge states. We now change the pump polarization from linear to circular and examine how the response of the edge modes is modified as we scan the energy of pump photons within the gap. In particular, we show the results for two pump choices: $\{h_+ = 0.004, h_- = 0\}$ and $\{h_+ = 0, h_- = 0.004\}$, and for three values of the pump momentum $k = 0.4\,\mathrm{K}$, $0.51\,\mathrm{K}$, and $0.6\,\mathrm{K}$. The left column in Fig. 3 shows the maxima of the $\psi_-$ amplitude on the left (red) and right (green) edges of the lattice vs $\varepsilon$. Note, that the Zeeman splitting and losses violate the time reversal symmetry and therefore changing just polarization of the pump from plus to minus does not results in a symmetric transformation of the resonant curves.

Because $\psi_+$ is generally weaker than $\psi_-$ for $\Omega > 0$ in the pump-free limit, the spin-positive pump most typically makes weaker resonances. To see this, compare maximal amplitudes achieved at the left edge for different pump polarizations, cf., red dots in Figs. 3(a,c,e) showing the spin-positive pump and in Figs. 3(b,d,f) showing the spin-negative pump. The same simple argument does not apply to the resonances corresponding to the modes on the right edge and one can see that the maxima of the green resonances in Fig. 3 are practically insensitive to the choice of either plus or minus polarized pump. We understand that this is related to the fact that for the states on the right edge the group velocity is opposite to the pump momentum, which impacts their excitation efficiency. Changing the pump momentum to negative $k < 0$, reverses the resonance peaks on the left and right edges in a symmetric fashion. An interesting situation is encountered at $k = 0.51\,\mathrm{K}$ when the left and right edge states have close energies, see Figs. 3(c) and 3(d). In this case by using different pump polarizations one can selectively excite the state on the right edge only, Fig. 3(c), or combination of the two strongly localized states with close amplitudes on the opposite edges, see Fig. 3(d).

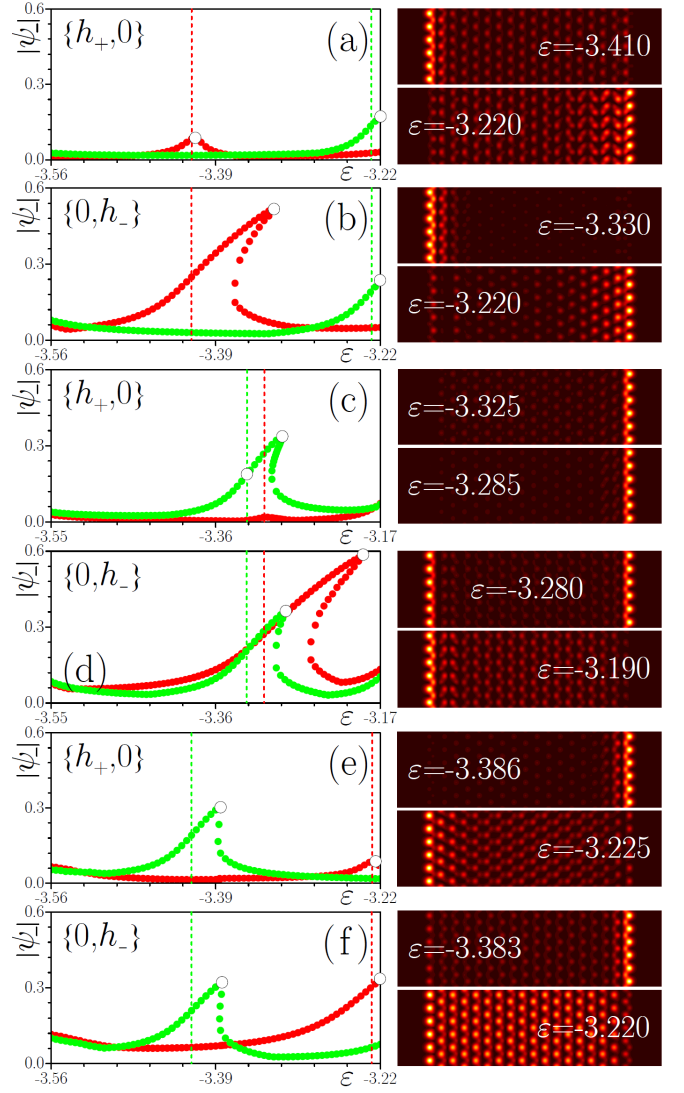

Fig. 3. (Color online) Left column shows amplitude of the spin-negative component on the left (red circles) and right (green circles) edges versus detuning for different input polarizations indicated in each panel: $k = 0.4\,\mathrm{K}$ in (a,b), $k = 0.51\,\mathrm{K}$ in (c,d), and $k = 0.6\,\mathrm{K}$ in (e,f). Distributions of $|\psi_-|$ in the right column correspond to the energies marked with open circles. Vertical red and green dashed lines mark energies of the linear pump-free edge states. The amplitude $h_\pm$ of the nonzero pump component is $0.004$ in all the cases.

We are returning to the case of linearly polarized pump $h_+ = h_-$, for stability analysis. We use direct numerical integration of Eqs. (1) with slightly perturbed inputs $\psi_\pm = u_\pm(1 + \alpha_\pm)e^{iky}$, where $\alpha_\pm(x,y)$ is the broadband $1\%$ noise and $u_\pm(x,y)$ is the edge state. Such inputs were allowed to evolve up to $t \sim 10^4$, which has allowed us to capture even weak instabilities and to accurately determine boundaries between stable and unstable edge states. Stable states are shown in black in Figs. 2(a) and 2(b), while the unstable ones are in red. The pump momentum in Fig. 2 is $k = 0.4\,\mathrm{K}$ and falls into the interval of the positive polariton mass. Performing similar simulations with the pump momentum shifted to the negative mass interval have not revealed a significant difference in the instability scenarios. Instabilities appear only for sufficiently large pump amplitudes close to the tip of the resonance. Within bistability domain solutions belonging to the upper branch are usually unstable. However, close to the point where middle and lower branches join a narrow detuning interval exists where solutions from the upper branch can be stable even within bistability domain (for $h_\pm = 0.004$ the width of this domain is $\delta\varepsilon \approx 0.006$). Outside the bistability domain the upper branch is always stable. Moreover, corresponding stable edge states can be very well localized, sometimes even better than states from the tip of the resonance curve, see Fig. 2(b). The low amplitude branches are always stable, while the middle branches are always unstable. If bistability is absent at low $h_\pm$, then the entire branch of solutions is usually stable. Typical dynamics of the instability development is shown in Fig. 4. The instability leads to the modulation of the polariton density along the edge and in the bulk of the lattice and to polarization rotation of the radiation pattern emerging inside the bulk, so that the alternating regions with dominating either spin-negative or spin-positive components propagate from the edge into the bulk. The instabilities reported here are likely to be associated with the complex underlying four-wave mixing processes and their classical and quantum properties require further investigation.

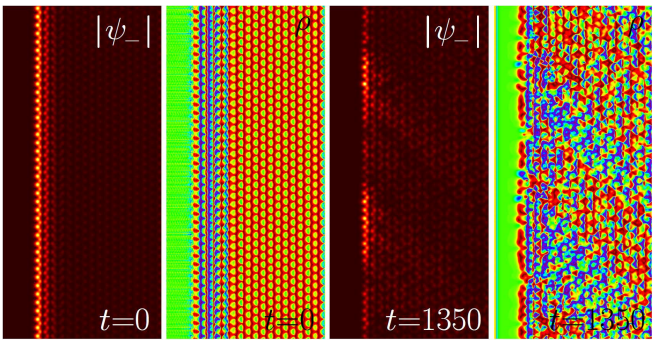

Fig. 4. (Color online) Instability induced dynamics of the edge state from the upper branch at $\varepsilon = -3.33$, $k = 0.4\,\mathrm{K}$, $h_\pm = 0.004$. $|\psi_-|$ and the local polarization degree $\rho$ are shown at the initial moment and in the advanced stage of the instability development.

Finally, we modeled an impact of the structural perturbations on dynamics of the edge states. We considered two types of perturbations, which can be encountered in practical devices. First one is the missing pillar and second one is random fluctuations of the energy resonances between different pillars associated with variations of the depth of the local trapping potential they create for polaritons. We have found that the propagation through the missing pillar results in some local reshaping of the polariton density accompanied by neither backscattering nor violation of the localization of the state around the lattice edge. In the case of random fluctuations of the energy resonances we have found that the fluctuations that are well below the energy value corresponding to the width of the topological gap preserve the edge states. However, the ones comparable and exceeding the gap width destroy the edge states and lead to the excitation of multiple modes in the bulk of the lattice. Numerical data demonstrating structural stability and topological protection of the edge states in both of the above cases are included in the Supplementary Materials [52].

In summary, we have proposed bistable polaritonic topological insulator, where external pump that compensates intrinsic losses in the microcavity allows to selectively excite desired modes, including well-localized topologically protected edge states. Bistability and instabilities of topological states can find their use in development of the topological quantum information processing schemes.

DVS acknowledges support from the Royal Society (IE 160465) and the ITMO University visiting professorship scheme via the Government of Russia (Grant 074-U01).